\documentclass{sig-alternate}
\usepackage{mathptmx} 

\usepackage{fancyhdr}
\usepackage[normalem]{ulem}
\usepackage[hyphens]{url}
\usepackage[sort,nocompress]{cite}
\usepackage[final]{microtype}
\usepackage[keeplastbox]{flushend}
\usepackage{xspace}
\usepackage{hhline}

\usepackage{amsmath}
\usepackage{bm}
\usepackage{tikz}
\def\binomialCoefficient#1#2{
  \the\numexpr 1\expandafter\bKN\expandafter{\the\numexpr #2\relax}{#1}\relax}
\def\bKN#1#2{
  \ifnum #1<0 *0
  \else    \ifnum 0<\numexpr 2*#1-(#2)\relax \expandafter\bKN\expandafter{\the\numexpr #2-(#1)\relax}{#2}%
           \else  \bkNK 1{#2}{#1}\fi\fi}
\def\bkNK#1#2#3{
  \ifnum #1>#3 
  \else  *(#2)/#1\expandafter\bNkKfi\expandafter{\the\numexpr#2-1\relax}{#1+1}{#3}%
  \fi}      
\def\bNkKfi#1#2#3#4{
 #4\expandafter\bkNK\expandafter{\the\numexpr#2\relax}{#1}{#3}}

\makeatletter
\newcommand*\etal[0]{{\it et~al}\@ifnextchar{.}{}{.\@\xspace}}
\makeatother

\newcommand\kB[0]{\,kB\xspace}

\usepackage[bookmarks=true,breaklinks=true,letterpaper=true,colorlinks,linkcolor=black,citecolor=blue,urlcolor=black,pdfusetitle]{hyperref}
 
\fancypagestyle{empty}
{
  
   \fancyhf{}
   \fancyfoot[C]{{This paper is distributed under the \href{https://creativecommons.org/licenses/by-nc-nd/4.0/}{CC BY-NC-ND} license}}
}

\title{Branch Predicting with Sparse Distributed Memories}

\author{Ilias Vougioukas\\
Arm Research\\
ilias.vougioukas@arm.com
\and
Andreas Sandberg\\
Arm Research\\
andreas.sandberg@arm.com
\and
Nikos Nikoleris\\
Arm Research\\
nikos.nikoleris@arm.com
}

\begin{document}
\maketitle
\thispagestyle{empty}
\pagestyle{plain}


\begin{abstract}

Modern processors rely heavily on speculation to keep the pipeline filled and consequently execute and commit instructions as close to maximum capacity as possible. To improve instruction level parallelism, the processor core needs to fetch and decode multiple instructions per cycle and has come to rely on incredibly accurate branch prediction. However, this comes at cost of the increased area and complexity which is needed for modern high accuracy branch predictors.

The key idea described in this work is to use hyperdimensional computing and sparse distributed memory principles to create a novel branch predictor that can deliver complex predictions for a fraction of the current area. Sparse distributed memories can store vast amounts of data in a compressed manner, theoretically enabling branch histories larger and more precise than the branch predictors used today to be stored with equal or smaller area footprint. Furthermore, as all the data is in a hashed format and due to the nature of the hashing scheme used, it is inherently harder to manipulate with known side channel attacks.

We describe our proof-of-concept and evaluate it against a state-of-the-art academic TAGE predictor. Our experiments are conducted on realistic synthetic branch predictor patterns and the Championship Branch Prediction traces and show competitive accuracy. Finally, we describe techniques that can be used to solve some of the challenges of processing with hyperdimensional vectors in order to deliver timely predictions.

\end{abstract}
\section{Introduction}
Over the past 15 years branch prediction has primarily focused on two designs, the Perceptron and tagged geometric history length (TAGE) predictors\cite{Jimenez,Seznec}. While the two predictors share some common insights that older predictors do not leverage, they accomplish their high predictions accuracy in very different ways.

These two designs improve on prior branch prediction technologies by making the observation that branch patterns of varied length occur. As such, long (global) histories can identify patterns that are elusive with shorter ones, conversely short histories are useful when longer histories fail to identify matching longer sequences. This made predictors more accurate but also lead to more complicated designs that take longer to deliver predictions\cite{Jimenez2000}.

\begin{figure}[!t]
\centering
\includegraphics[width=.85\columnwidth]{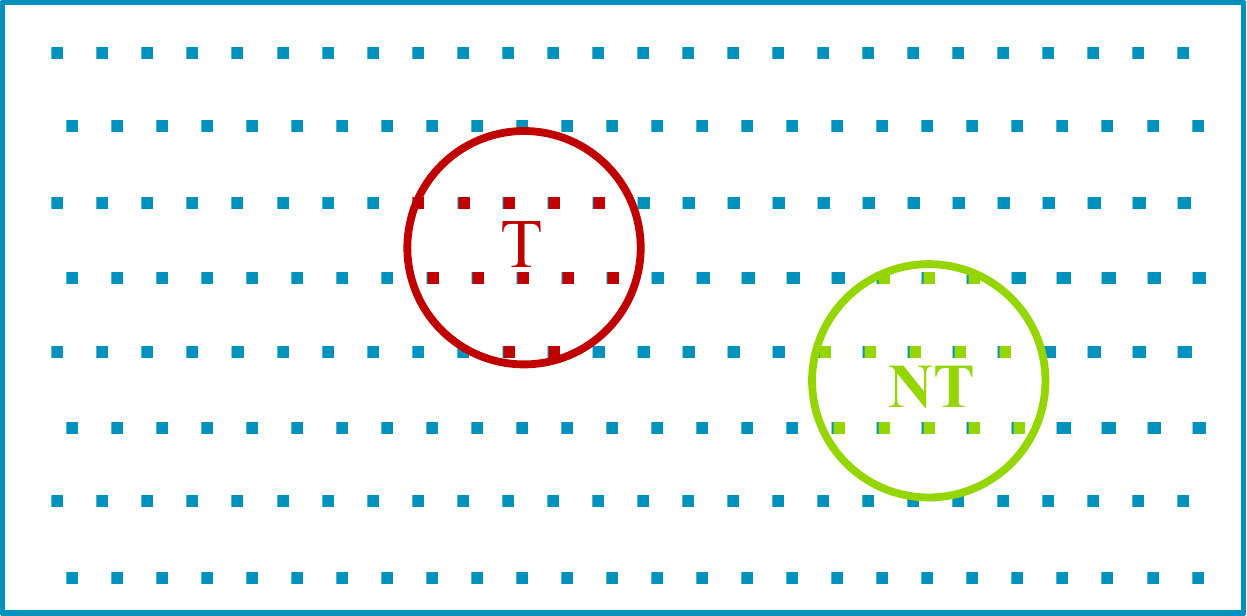}
\caption{HYPRE operates in a high dimensional space. Through training the taken (T) and not taken (NT) points can be easily classified.}
\label{motivate}
\vspace{-.6cm}
\end{figure}

Beyond that similarity, their approach at capturing patterns is vastly different. TAGE predictors, deploy a straight-forward approach where the geometric histories are stored in tagged tables. These tagged geometric history tables comprise over 85\% of the entire design \cite{Seznec}.

Contrary to the aforementioned, the state-of-the-art version of Perceptron \cite{Jimenez} does not access a distinct prediction from a table. Instead it tracks certain features from a number of tables, among them the global history (similar to TAGE), and sums up their weights. Depending on whether the sum is above or below a threshold, the predictor decides whether to predict the branch as taken or not taken.

These fundamentally different ways of calculating the prediction, also gives the two predictors distinct attributes that affect their behavior. As Perceptron predictors produce their outcome from weights used by multiple entries the design is, in theory, able to store patterns and certain important features in a more dense manner. This compressed way of storing however, makes it difficult to pin-point exactly what affects a specific prediction.

To juxtapose this, the distinct tables in TAGE that store entries separately provides clear tractability for each prediction, making the overall reasoning around the design much easier. This likewise causes a trade-off in terms of area efficiency, which depending on the characteristics of the workloads can under-utilize entire tables that don't match the pattern lengths.

Overall both TAGE and Perceptron, deliver very accurate predictions, questioning whether there is any room for improvement. Recent studies however, have shown that while these predictors are highly accurate, they still miss branches classified as hard-to-predict. These branches which were until recently considered unobtainable have been shown to have significant performance impact\cite{Chauhan2020}. More pertinently though, the studies show that branch prediction still has room for improvement, even in terms of accuracy \cite{Lin2019}.

In addition to this, recently discovered side-channel attacks \cite{Kocher,Canella2019,Evtyushkin2018,Weisse2018,Koruyeh2018} that target the branch predictor have been shown to be difficult to mitigate without a penalty in performance, area or complexity \cite{Vougioukas2019, Kiriansky2018,Yan2018,Qureshi2018,McIlroy2019,Khasawneh2018}.

This motivates our work to investigate a different approach that can lead to improvements in one of the important factors that govern branch predictor design; namely accuracy, area, speed of prediction, and security. With the above in mind, our contributions in this work are:
\begin{itemize}
    \item We introduce a novel branch predictor design family inspired by hyper-dimensional computing and sparse distributed memories called HYPRE.
    \item We show that using ``hypervectors'', HYPRE uses significantly less space to store history patterns both long and short. We show how HYPRE can be implemented in hardware and identify all the potential difficulties in implementation
    \item We detail how the storing mechanism for HYPRE is not susceptible to most Spectre type attacks, as it is more hardy to branch prediction poisoning.
    \item We evaluate how HYPRE compares to TAGE, Perceptron and bimodal.
\end{itemize}
\section{Background}

\begin{figure}[!t]
    \centering
    \includegraphics[width=\columnwidth]{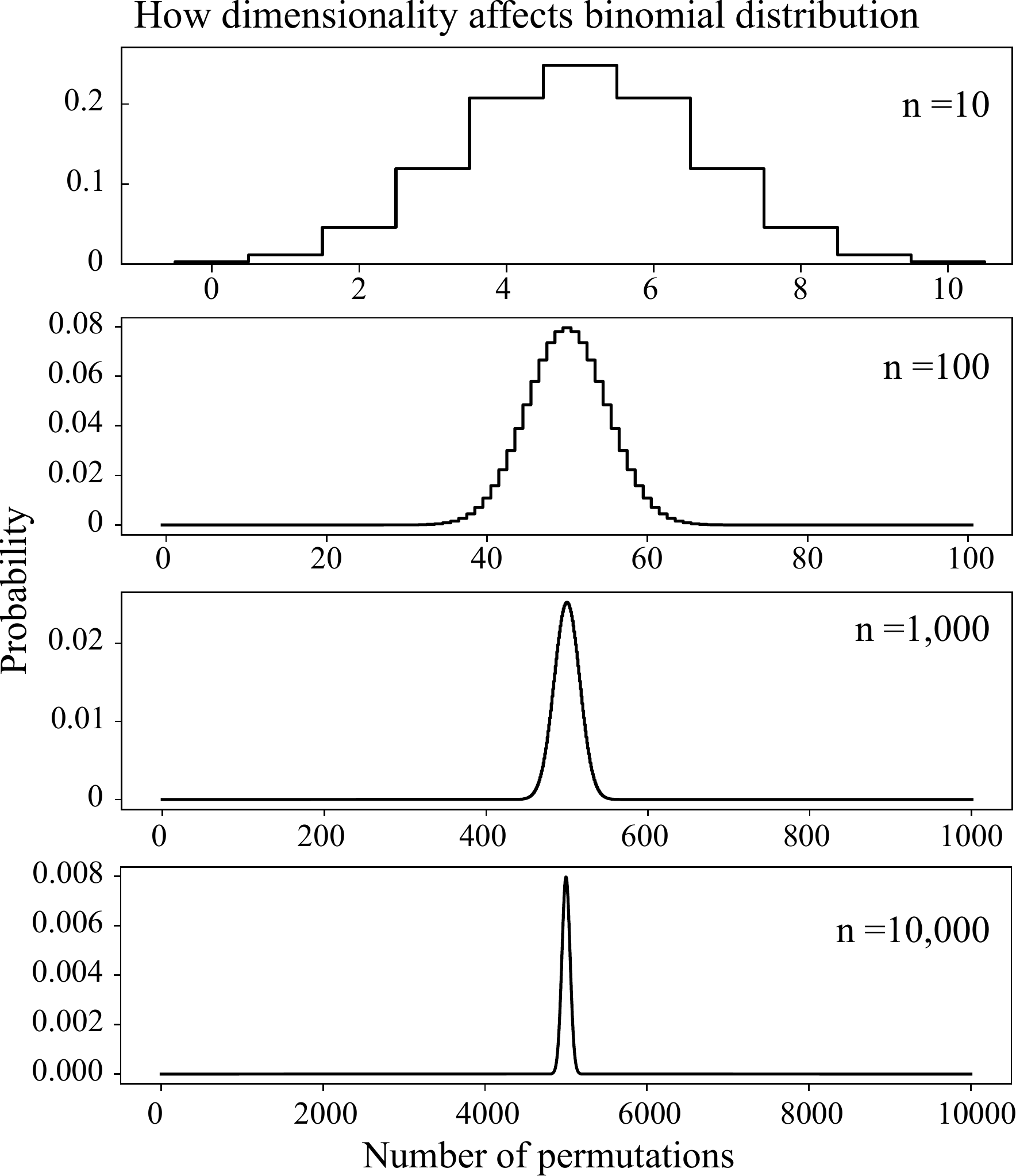}
    \caption{Higher dimensionality means that the probability distribution is tighter. Two random hypervectors of 10\,000 bits are practically guaranteed to mismatch in roughly 50\% of their bits.}
    \label{Binom}
    \vspace{-.3cm}
\end{figure}

\subsection{Branch prediction design}
Branch prediction over the years has moved from small and simple designs\cite{Evers1996,Yeh1991} to complex structures storing long histories. We briefly outline the two most popular designs used today: the TAGE \cite{Seznec}, and the Multiperspective Perceptron predictors \cite{Jimenez}.

\subsubsection*{TAGE-based predictors}
The TAGE predictor is one of the most accurate designs \cite{Seznec, Vougioukas2019} in use today. TAGE uses global history of varying length to index into its prediction tables which are tagged using branch PC bits. The history lengths form a geometric series, hence the name tagged geometric (TAGE). The matching prediction from the table with the longest history length is used to predict a branch's outcome. If no match is found it uses its base predictor, a bimodal design, as a fall-back mechanism \cite{Seznec2006}.

While the TAGE is very accurate and reliable, its tables require a substantial amount of space. Furthermore, the medium length history tables tend to be underutilized, which suggests that there is still potential to reduce the size of the predictor and compress the information stored in the sparsely used tables.


\subsubsection*{Perceptron type predictors}
Another popular design that is in widespread use today is the Perceptron predictor \cite{Jimeneza, Seznec2005}. Loosely based on neural network theory, perceptron-type predictors achieve high accuracy from efficiently stored state.

The principle behind the state-of-the-art Perceptron \cite{Jimenez}
uses tables with sets of weights that are accessed using a set of useful features. The output and confidence of the predictor depends on the sign of the sum of all the feature$\times$weight products. The confidence can be deduced from the magnitude of the value. Modern versions of Perceptron have reduced the amount of calculations required for each prediction and use multiple hashing tables of weights that are indexed using both the history and the branch address \cite{Tarjan2005}. These tables are commonly referred to as feature tables \cite{Jimenez}. As the entries of each feature table are used for multiple predictions and then summed, it is very difficult to do topical changes to the weights to better track the behavior of a single branch pattern.

Furthermore both TAGE and Perceptron have been shown to be susceptible to Spectre-type side-channel attacks. Entries in TAGE tables can be isolated and overwritten, poisoning the useful counters and potentially flipping the prediction. Perceptron weights can also be manipulated as each weight entry is used in multiple predictions \cite{Kocher,Canella2019,Evtyushkin2018,Weisse2018,Koruyeh2018,Vougioukas2019}.

\subsection{Sparse Distributed Memories}
\subsubsection*{Hyper-dimensional Computing Principles}

\begin{figure}[!t]
    \centering
    \begin{tikzpicture}[rotate=180]
        \foreach \n in {1,...,10} {
            \foreach \k in {0,...,\n} {
                \node at (\k/1.25-\n/2.5,\n/3) {$\binomialCoefficient{\n}{\k}$};
            }
        }
    \end{tikzpicture}
    \caption{Pascal's triangle  for dimensions n between 1 and 10 . This shows how many bits are expected to match between two random hypervectors. The higher the dimensionality, the larger the probability that two randomly generated vectors have roughly half of their bits match.}
\label{pascal}
\end{figure}
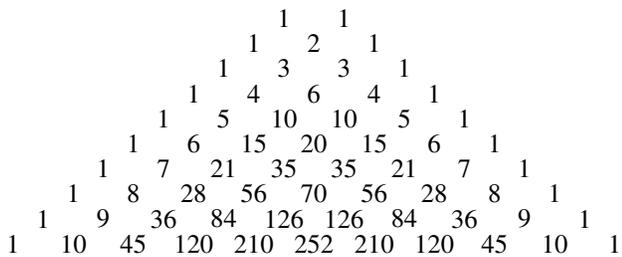

The theory of Sparse Distributed Memories (SDM) is at the core of hyper-dimensional computing (HDC)\cite{kanerva1988sparse}. The main idea behind it is that data can be stored in a content addressable way in very large binary vectors, usually on the order of thousands of elements called HyperVectors (HV). In this expanded form, all stored objects have a distinct combination of values for their dimensions, similar to a unique hash. The probability and cumulative distribution function that the bits of two random vectors match follows a binomial distribution:

\begin{equation*}
P(k)= \frac{\binom{n}{k}}{2^n}, \qquad F(k)=P(X<k)=\frac{\sum_{n=0}^{\lfloor k\rfloor}\binom{n}{k}}{2^n}
\end{equation*}
In the above equations $n$ is the number of dimensions and $k$ is the number of bits that are different. What follows from the formula, is that as the dimensionality of the vectors increases (e.g., more than 1\,000) the probability that two randomly chosen vectors are the same becomes infinitesimally small. To be precise, approximately half of the bits of the two vectors are the same and half are different (\autoref{Binom}). Conversely, \emph{if two vectors match more than 50\% then they are not random but correlated with near certainty.}

Vectors can be manipulated, allowing the stored data to be directly processed in high dimensional space. These vectors are traditionally selected to operate in the discrete binary set of  \{-1, 1\}. However, as 0 and 1 is the common notation for binary logic, all algebraic operations have been converted to their equivalents using the \{0, 1\} set.

In the scope of this work, we focus on the three most important operations:
\begin{itemize}
    \item Addition: To correlate two or more vectors the only necessary operation is an element-wise addition. In the binary form, this element wise addition is effectively a majority vote, which randomizes the output when there are equal number of 0s and 1s, as shown in \autoref{Examples}. Adding allows independent HVs to create a new, composite HV which strongly correlates to all the original HVs used in the summation.
    \item Multiplication: HVs can also be XORed to create new vectors that represent an entanglement of the initial vectors. The multiplication operation is distributive over addition.
    \item Rotation: This operation can be used to do an element-wise circular shift on all the bits in a vector. This operation is distributive over both addition and multiplication. 
\end{itemize}
It is also worth noting that the above operations are reversible.

\begin{figure}

\begin{align*}
&\text{Addition:}&\qquad\quad  &\begin{bmatrix} 0 & 1 & 0 & 0 & \dotsm & 1 & 1 & 0 & 1 \end{bmatrix}& \\ \nonumber
&                &       &{\fontsize{7.5}{7} \begin{matrix} + & + & + & + & \dotsm & + & + & + & +\end{matrix}}&\\ \nonumber
&                &       &\begin{bmatrix} 0 & 1 & 1 & 1 & \dotsm & 1 & 0 & 1 & 1 \end{bmatrix}&\ = \\ \nonumber
&                &       &\begin{bmatrix} 0 & 1 & \textbf{0} & \textbf{1} & \dotsm & 1 & \textbf{1} & \textbf{1} & 1 \end{bmatrix} \nonumber
\end{align*}

\begin{align*}
&\text{Multiplication:}&\quad &\begin{bmatrix} 0 & 1 & 0 & 0 & \dotsm & 1 & 1 & 0 & 1 \end{bmatrix}& \\\nonumber
&           &            &{\fontsize{7.3}{7} \begin{matrix}\oplus & \oplus & \oplus & \oplus & \dotsm & \oplus & \oplus & \oplus & \oplus\ \,\end{matrix}}&\\ \nonumber
&           &            &\begin{bmatrix} 0 & 1 & 1 & 1 & \dotsm & 1 & 0 & 1 & 1 \end{bmatrix}&\ = \\ \nonumber
&           &            &\begin{bmatrix} 0 & 0 & 1 & 1 & \dotsm & 0 & 1 & 1 & 0 \end{bmatrix}& \nonumber
\end{align*}

\begin{align*}
&\text{Rotate:}&\qquad\quad\;\;&\begin{bmatrix} 0 & 1 & 0 & 0 & \dotsm & 1 & 1 & 0 & 1 \end{bmatrix}& \\ \nonumber
&              &       &{\fontsize{6.5}{7} \begin{matrix} \ll & \ll & \ll & \ll & \dotsm & \ll & \ll & \ll & \ll\ \,\end{matrix}}&\\\nonumber
&              &       &\begin{bmatrix} 1 & 0 & 0 & 1 & \dotsm & 1 & 0 & 1 & 0 \end{bmatrix}&\nonumber
\end{align*}
    \caption{The defined HV elementwise operations. Addition works as a  majority, multiplication as an XOR and Rotate as a circular shift operation.}
    \label{Examples}
    \vspace{-.3cm}
\end{figure}

\subsubsection*{Sparse Space Representation}
\label{sparsespace}
As the space exponentially increases when vectors have many dimensions, sparse memories can only use a few discrete (physical) locations to store information efficiently. This can be seen in \autoref{motivate}, where a 2-dimensional representation is shown for easier visualization. In the figure, the box represents the entire space while the blue dots the discrete, physical locations. The rest of the space is composed notionally by interpolating from the nearby discrete locations. To achieve this, data is stored in a distributed way that stores a vector across a region in the hyper-dimensional plane.

This distributed storage works because the memory only needs a few of the discrete locations to store or retrieve information. The entirety of the space is constructed by performing a write to all the locations in a notional radius to store and by sampling and adding them to read. The rationale for the name is now obvious: the memory is sparse because the physical locations are an infinitesimally small subset of the memory space; it is distributed because a pattern is stored in many locations and retrieved by statistical reconstruction from many locations. This can be seen in \autoref{motivate} where Taken (T) and Not Taken (NT) objects are not centered on a discrete location but are instead composed by averaging out all the discrete locations within their respective radii.
\begin{figure}[!t]
    \centering
    \includegraphics[width=\columnwidth]{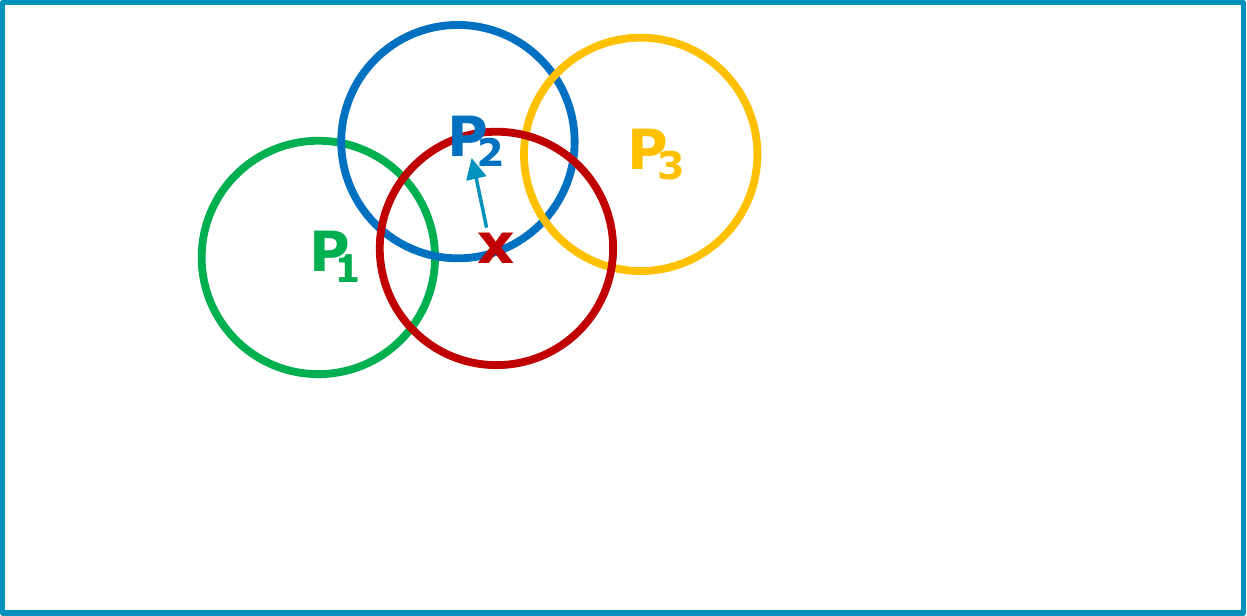}
    \caption{Sparse distributed memories can find the best fitting data even in the presence of noise or nearby data with high accuracy. In this case the X with converge to P2.}
    \label{find}
    \vspace{-.5cm}
\end{figure}

Distribution enables this content addressable  memory to retrieve a stored pattern even when the input cue only partially matches any stored pattern. In \autoref{find}, we see that while points P1, P2 and P3 are overlapping, the query X point converges to P2. This happens because during retrieval all of the locations within the radius of X are summed up and it also renders the memory robust in case of failure of portions of the addressing or storage hardware. The Taken (T) and Not Taken (NT) objects are not centered on a discrete location but are instead composed by averaging out. A benefit of this approach is that when trying to retrieve a specific piece of information from a memory, one lookup is enough to determine whether it has been stored or not.

The above observation also points to the reconstructive properties of SDMs and their ability to deliver accurate results even in the presence of noise. For example, using 1\,000-bit vectors, gives rise to a space of 21\,000 possible patterns. In this space, 1/1000 of the patterns are within 451 bits of any given pattern, and all but 1/1000 of the patterns are within 549 bits. The extremely large number of patterns that are so close (±49 bits) to the mean distance of 500 bits between two random patterns is crucial to the memory’s ability to make connections between patterns that seemingly have little to do with each other. This is an inherent property as the larger the dimensionality of the vectors the smaller the chance that two vectors are similar. In fact, for vectors of 10\,000 dimensions the probability that at least 47\% of their bits are different is 99.9999\%. With tighter similarity constraints the probability that two random vectors can be mistaken is virtually zero.

The above property is particularly important in the case of speculation and branch prediction as it makes the design robust against most Spectre type attacks. In traditional predictors, an attacker can target an entry and change the prediction by figuring out the index patterns of the memories they are stored in and force the predicted value to change through aliasing. This cannot be leveraged however, in an HDC based predictor because of all the information is stored in one very large vector.  Furthermore,branch patterns are translated to arbitrary HyperVectors making it impractical to manipulate the values in such a way that  can alter the vector containing all the pattern information. This means that a branch predictor based on HDC can never alias. In fact, the address bit pattern is expanded multiple times and can not therefore be brute forced as it would take many orders of magnitude more time than is realistically available to perform the attack.
\vspace{-.2cm}
\section{HD Branch Prediction}
 \begin{figure}[!t]
    \centering
    \includegraphics[width=\columnwidth]{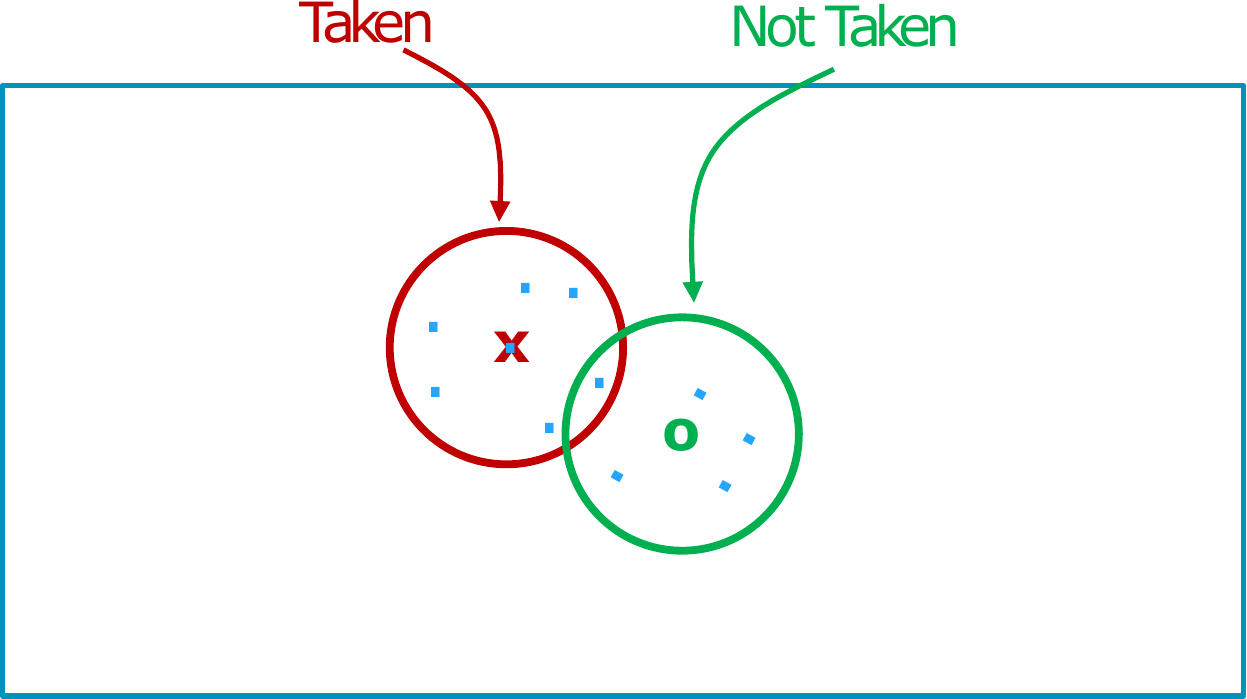}
    \caption{The Taken and Not-Taken Vectors in HD representation. The Vectors are a sum of only the patterns they have trained on.}
    \label{TNT-b}
    \vspace{-.3cm}
\end{figure}
In this section we will describe HYper-dimensional PREdictor (HYPRE), a design that uses HDC to deliver fast and accurate predictions, and discuss certain important design considerations that this approach requires.

\subsubsection*{How do HVs capture branch patterns}

In order to understand how HYPRE functions, it is important to explain how branches and branch patterns are converted into HVs. To do this the HD engine assigns a random HV to each branch instruction. This can be done by picking a random sequence of zeros and ones when a branch is first encountered, and subsequently store it into a structure that serves as a ``dictionary'' that maps a program counter (PC) to the assigned HV. To do the assignment efficiently, it is possible to use a hashing mechanism that processes the branch PC address directly into HVs. This saves space but requires the hash to be processed every time the mapping is needed.

To compose sequences of branches, we use the mathematical operations described in \autoref{Examples}. Assuming a sequence of branch PCs the unique HV that describes it can be created as shown below:
\begin{equation*}
    PC_A \rightarrow PC_B \rightarrow PC_C \Longleftrightarrow (HV_A \ll 2) \oplus (HV_B \ll 1) \oplus HV_C
\end{equation*}

The HVs are rotated so that the pattern can capture their position in the sequence while the multiplications merge their respective results into one HV that uniquely identifies this specific pattern. For long histories, this process can be slow as thousands of rotations and XOR calculations are needed to get the final HV. This can be dramatically sped up using two observations.

First, when using a branch pattern of fixed length for each consequent prediction, only the oldest PC is discarded while a new one is added at the top of the sequence as show below:
\begin{align*}
&    PC_A \rightarrow PC_B \rightarrow PC_C \Rightarrow \underbrace{(HV_A \ll 2) \oplus (HV_B \ll 1) \oplus HV_C}_{HV_{ABC}}\\
\\
&    PC_B \rightarrow PC_C \rightarrow PC_D \Rightarrow (HV_B \ll 2) \oplus (HV_C \ll 1) \oplus HV_D\\
\\
&\Rightarrow (HV_A \ll 2) \oplus (\bm{HV}_{\bm{ABC}} \ll 1) \oplus HV_D
\end{align*}
As the operations are reversible, rotating $HV_A$ by 2 and XORing it removes the oldest branch while the only thing that is left to do is to rotate once and combine it with the new $HV_D$. This means that once the main chain has been calculated, 4 operations are only needed to fetch the next HV pattern regardless of the length of the branch history.

The second observation derives from the fact that all but the last multiplication can be done off the critical path. This means that at every prediction the partially calculated sequence is ready and one XOR operation factors in to the timeliness of the prediction.

Another simpler way of creating the branch patterns is to just keep track global history and use the bits along with the PC as the input to the HV mapping engine to create a unique HV for that combination. Expanding on that idea it is possible to incorporate more features similar to multiperspective perceptron\cite{Jimenez} to improve the accuracy.

\subsubsection*{Predict: How predictions are made}
Now that the generation of HV patterns has been explained, we focus on the underlying mechanism that ultimately decides whether a branch should be taken or not. To simplify, we assume a predictor that only uses a branch sequence to create an HV like described above. When a branch arrives and the HV for the pattern is formulated we perform a ``look-up'' operation in the SDM. However, creating the entire memory out of all of the discrete locations as described in \autoref{sparsespace} is impractical.

To solve this problem we use an optimization trick based on the fact that the only part of the memory that is useful is the one that describes the Taken (or Not Taken) outcome (Figure \ref{TNT-b}). In \autoref{sparsespace}, we described how to recall  and store from a given point in memory, all the hard locations within a radius need to be summed added up. Instead of this, it is possible to do the inverse meaning, to have one HV that is already the summation of all the discrete locations that contribute to it. This effectively is the Taken HV that the input is compared with. The comparison finds the hamming distance between the two vectors. This can be done with an element wise XOR followed by a population count. 

The determination of whether the two vectors are a match is based on a threshold that derives from the binomial distribution characteristics mentioned in the background section and depends on the dimensions of the vectors. Finally if the incoming HV is matches the Taken HV above the threshold then the branch is Taken. Otherwise if the bits that match are below the threshold the outcome is Not Taken. Designs can use both a Taken and a Not Taken vector which allows more introspection, but it is not necessary as the the two outcomes are by definition mutually exclusive.
\begin{figure}[!t]
    \centering
    \includegraphics[width=\columnwidth]{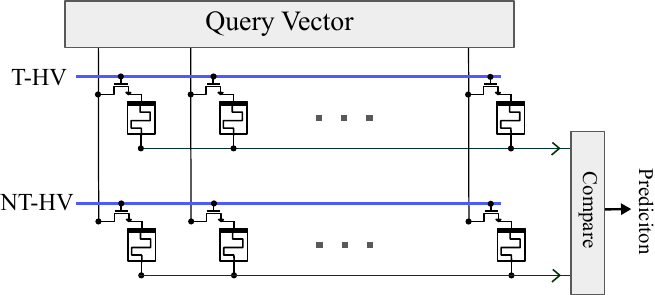}
    \caption{The analog comparator circuitry can enable prediction on time. The bits that match between the prediction query and the Taken/Not Taken vectors allow current to flow. If the accumulated current overcomes a threshold then there is a match.}
    \label{analog}
\end{figure}

One noteworthy design aspect of the matching mechanism is that the population count is an complex and expensive calculation that does not scale well with vector size. As the vectors in sparse memories are thousands of bits long, this makes a digital pop count approach impractical as it would be too slow to deliver a prediction result in time. For this reason, many HDC studies have used an analog equivalent to perform the comparison and the pop count in one step \cite{Rahimi2017,Imani2019,Imani2020,Karunaratne2019}. This is also depicted in \autoref{analog} where the incoming query vector is compared with a stored taken (T-HV) and a not take vector (NT-HV). The elements that match allow current to flow which accumulates on the line and is sensed and compared to the threshold. If both pass then the highest one is selected for the prediction.
\subsubsection*{Update: How the predictor is trained}
To train the predictor the only thing that is necessary is to add the query vector to the Taken or Not Taken vector depending on how the branch has resolved. As SDMs have the ability to perform one-shot learning \cite{Burrello2018}, \emph{correctly classified patterns do not need to be reinforced into the model}. Instead, HV values are added when the prediction showed marginal correlation or that the vectors were independent. When a misprediction occurs, HVs can be subtracted from from the vector that incorrectly matched the pattern. This effectively removes it from the set and ensures that it will not cause a wrong prediction again.

For better accuracy, the addition and subtraction uses more than one bit. The extra bits do not leverage the fidelity that magnitude in each dimension introduces to improve the accuracy. As the comparison is very time sensitive, a complex mechanism to provide a better comparison is likely to critically delay the prediction. Instead the added bits per element help the counter not ``lose'' information when multiple additions or subtractions occur as each dimension has more states to occupy this way. The comparison remains simple as only the sign bit is used to determine whether the dimension is a 0 or a 1. \emph{The number of extra bits in the storage vectors is called saturation.}
\begin{figure}[!t]
    \centering
    \includegraphics[width=\columnwidth]{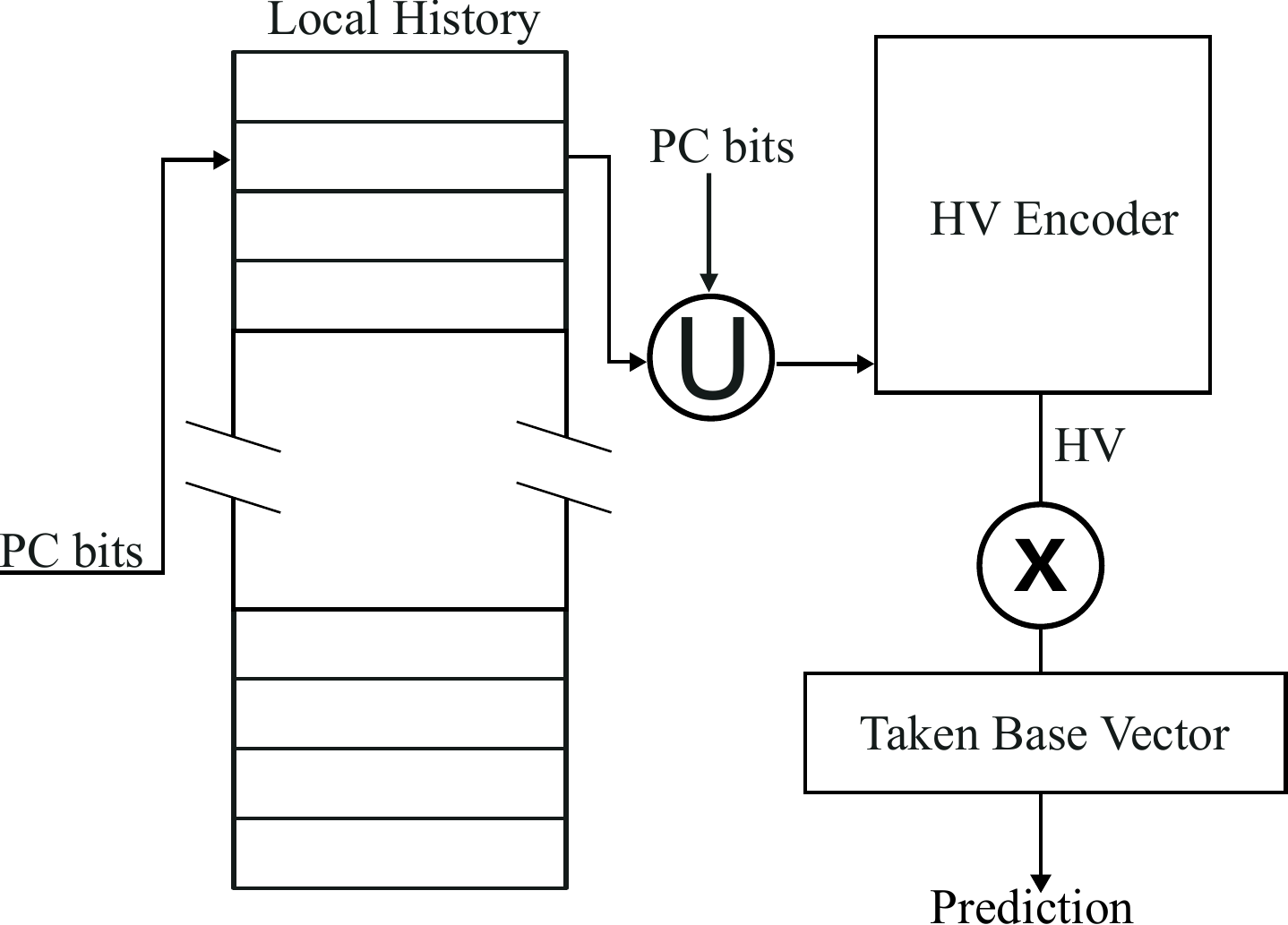}
    \caption{The HD-Bimodal design. The $\bigotimes$ symbolizes the comparator between the query and the stored T vector, while the \raisebox{.5pt}{\textcircled{\raisebox{-.9pt} {\small U}}} operator concatenates the output from the local history table with the PC.}
    \label{HD-BM}
\end{figure}
\subsection{The HYPRE design}
We propose the HYperdimensional PREdictor (HYPRE) prototype to evaluate the above mechanism as a branch predictor with potential to achieve high accuracy. We outline the high level design used in our experiments and note some practical fine tuning to improve its performance.
\subsubsection*{The HD base predictor}
\begin{figure}[!t]
    \centering
    \includegraphics[width=.82\columnwidth]{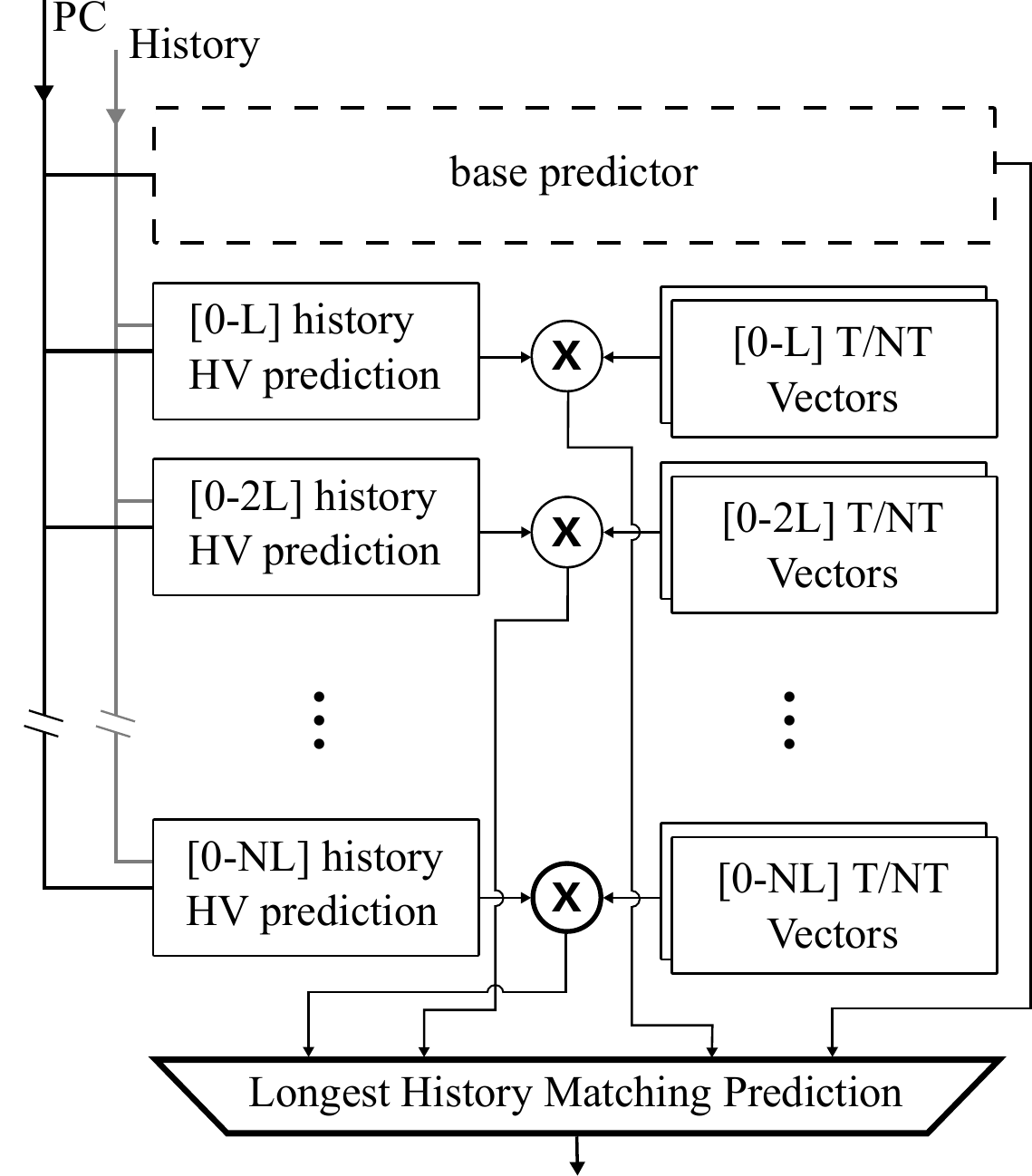}
    \caption{The outline of the HYPRE predictor. The $\bigotimes$ symbolizes the comparator between the query and the stored T/NT vectors. The vector that matches with the longest history is picked. If no vectors math it uses the base predictor}
    \vspace{-.3cm}
    \label{HD-BP}
\end{figure}

Similar to TAGE and Perceptron, HYPRE also relies on progressive geometric histories to deliver accurate predictions. This can be seen in \autoref{HD-BP} where the predictor generates multiple query HVs, one for each history length. These HVs are then compared to both the stored Taken and Not Taken vectors that correspond to their respective history length. The reason both are used is because it provides helps identify patterns that correlate with both sets.

The queries that passed the threshold and therefore matched are forwarded for to be compared based on their history length where the longest pattern is selected. If none of the vectors are confident that the pattern has been observed previously the design uses a fallback base predictor which is configure as a standard bimodal predictor.

Training is performed on all the vectors that either mispredict or have marginal confidence in their prediction. In some pathological cases a mispredict can lead to over-correcting. This can happen, for instance, when a common short pattern repeats as part of a longer sequence. After the the longer sequence is mispredicted for the first time, an addition sequence mispredictions will occur due to the fact that through training the shorter history length vectors perform a subtraction in the vector that delivered the short term prediction followed by an addition to the vector covering the opposite result. For example if the Taken vector for a small 3-branch pattern of only taken is corrected because a longer pattern emerges, the T,T,T sequence will be subtracted from that vector and added to the Not Taken vector which could be empty as no events so far required an insertion into its set. This temporarily drops the performance.

\begin{table}[!t]
\centering
\begin{tabular}{c|c|c||c||c}
Vector Size & History & Entries & Accuracy  & Size    \\ \hhline{=|=|=||=||=}
128\,b               & 2\,b                 & 1024              & 72.30\%      & 3\,kb  \\
1024\,b               & 2\,b                 & 1024              & 79.23\%   & 4\,kb  \\
2048\,b               & 2\,b                 & 1024              & 79.23\%   & 8\,kb  \\
1024\,b               & 2\,b                 & 2048              & 79.23\%   & 6\,kb  \\
1024\,b               & 3\,b                 & 2024              & 85.37\%   & 3\,kb  \\
1024\,b               & 4\,b                 & 512               & 89.35\%   & 4\,kb  \\
1024\,b               & 4\,b                 & 2024              & 90.16\%   & 10\,kb \\
1024\,b               & 6\,b                 & 1024              & 91.24\%   & 8\,kb  \\
1024\,b               & 6\,b                 & 2048              & 92.73\%   & 14\,kb \\
1024\,b               & 8\,b                 & 1024              & 93.41\%   & 10\,kb
\end{tabular}
\caption{HD bimodal configurations ordered by accuracy. Assuming that the hypervector is large enough to guarantee randomness, the predictor needs enough entries to not suffer from significant aliasing. Beyond that point the design is mostly sensitive to local history.}
\label{BM-Spec}
\end{table}

\begin{figure}[t]
    \centering
    \includegraphics[width=\columnwidth]{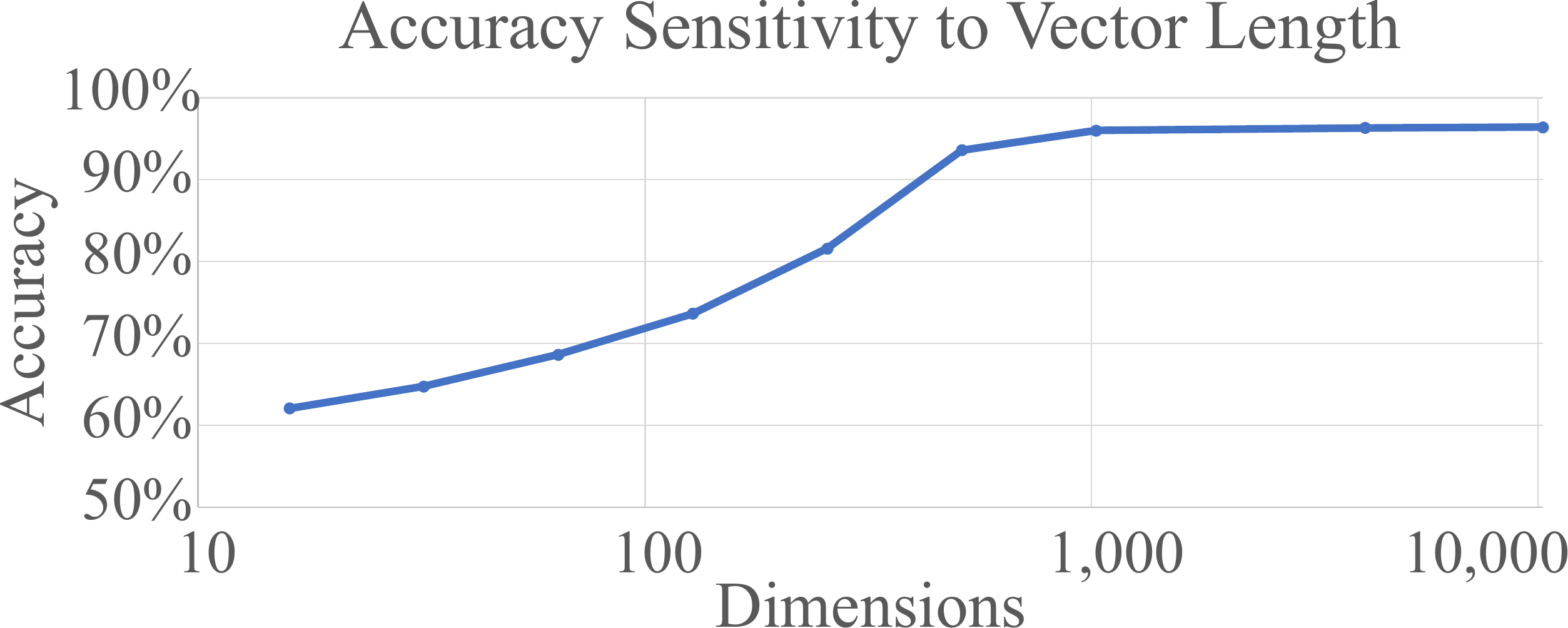}
    \caption{The vector length only affects the efficacy of the predictor when the length is too short. After 1024 bits the accuracy plateaus.}
    \label{sensei}
\end{figure}

This can be amended in two ways. First, when mispredicting instead of a full removal from a vector the system can perform a randomization of some of the bits to partially erase the prediction but not completely eliminate it. This makes sure that the most common pattern does not get removed completely from the set.

Second, when storing a pattern into a vector, deliberately randomize a small fraction of its elements to reduce its certainty. This does not affect repeated patterns as they will reach full certainty over time, but removes the absolute certainty of the one off event. This approach does the opposite of the first proposal; it makes sure that no insertion will ever register with 100\% confidence so that more frequent cases are preferred, but with enough confidence to correlate if no other option has been stored.

\subsubsection*{Multi-History HYPRE with HD-fallback}
Another area which optimization is possible is to improve the base predictor performance by introducing HDC aspects to it. The bimodal predictor delivers a prediction based on a counter using 2 local history bits. The outcome depends on whether the most significant bit (MSB) from the counter is 1 or 0 for taken and not taken predictions.
\begin{table*}[!ht]
\centering
\begin{tabular}{lcc}
 & Ideal & Realistic \\ \hline
\multicolumn{1}{|l|}{\begin{tabular}[c]{@{}l@{}}History Tables\end{tabular}} & \multicolumn{2}{c|}{8} \\ \hline
\multicolumn{1}{|l|}{History Length} & \multicolumn{2}{c|}{8, 32, 128, 256, 512, 1024, 2048, 4094} \\ \hline
\multicolumn{1}{|l|}{Vector Length} & \multicolumn{1}{c|}{4096\,b} & \multicolumn{1}{c|}{\begin{tabular}[c]{@{}c@{}}{[}8--256{]}: 1024\,b\\ {[}512--4096{]} 4096\,b\end{tabular}} \\ \hline
\multicolumn{1}{|l|}{\begin{tabular}[c]{@{}l@{}}%
        Base predictor
    \end{tabular}} & %
    \multicolumn{2}{c|}{%
        \begin{tabular}[c]{@{}c@{}}%
            HD base predictor: vector length 1024 bit, 4 bit local history, 2048 entries%
        \end{tabular}} \\ \hline
\multicolumn{1}{|l||}{\textbf{Total size}} & %
    \multicolumn{1}{l||}{%
        \begin{tabular}[c]{@{}l@{}}%
            4096\,b x 4 bits x 8 x 2 T/NT vectors +\\%
            8 query vectors x 4096\,b +\\%
            40\,b x 4094 entries +\\%
            10\,240 base predictor bits\\%
            = 468\,912\,b (\textbf{58.614\kB})%
        \end{tabular}%
    } & \multicolumn{1}{l|}{%
        \begin{tabular}[c]{@{}l@{}}%
            1024\,b x 4\,b x 4 x 1 T vector +\\%
            4096\,b x 4\,b x 4 x 1 T vector +\\%
            4 query vectors x 1024\,b +\\%
            4 query vectors x 4096\,b +\\%
            1\,b x 4094 entries +\\%
            10\,240 base predictor bits \\%
            = 116\,734\,b (\textbf{14.59\kB})%
        \end{tabular}} \\ \hline
\end{tabular}
\caption{The configuration used in the experiments. The 58.614kB ideal version is the one used for the analysis in the remainder of this paper.}
\label{HD-Table}
\end{table*}
To augment this design, we propose a table indexed by the PC. This outputs local history bits that in turn, are used along side the branch PC to generate an HV for the base predictor. This design is shown in \autoref{HD-BM} and has certain advantages over the standard bimodal. First, in the original bimodal the predictions rely solely on the MSB of the selected entry. This variant allows for all the bits to be reconfigured dynamically per PC to deliver a more accurate outcome.

Second, it allows for more local histories to be used to identify more complex patterns. This can help with cases which are pathological to bimodal such as oscillations between taken and not taken for a specific branch PC.

\section{Experimental Setup}
To evaluate HYPRE we perform a set of limit and sensitivity studies on an ideal proof-of-concept version of the predictor with all of the optimizations. Before doing this, a sensitivity study is conducted using synthetic traces, in order to find the correct configuration for the base predictor. The synthetic traces are comprised of a mixture of easy and hard to predict patterns, deliberately selected to stress all the tables.
\subsection{Base predictor configuration}
\autoref{BM-Spec} shows variations of the design that change the vector length, the number of bits used to store local history, as well as the amount of entries that the local history table will have. From the results we can make the following observations. The first observation is that when vector size is small, the HV cannot store all of the branch patterns without saturating, and as a consequence the accuracy of the predictor is very low. As shown in \autoref{sensei}, increasing the vector length causes the notional space grow exponentially and dramatically reduces the chance that the HV patterns interfere with one another. Beyond 1024 bits the accuracy experiences diminishing returns.

The second observation is that while intuitively we expect the entries into the local history table to heavily affect the accuracy of the predictor, the analysis points to the opposite. In fact, even with some aliasing in the Local History tables the predictor is able to retain its accuracy. This happens as the local history bits are concatenated with the PC bits to generate the unique HVs despite the shared local history entries. In our final design we opt to use 2048 entries as reasonable amount of entries.

The third observation is that the number local history bits per entry are the most important factor that affects the accuracy of the hyper-dimensional base predictor. Using two bits delivers 79\% accuracy, which is close to the reference bimodal (10\,kb with hysteresis) accuracy of 82\% shown in  \autoref{BM-Compare}. As the local history bits are increased so does the overall accuracy of the predictor, where beyond 8 bits diminishing returns start kicking in and a much larger local history is required. To keep to a similar budget of 10\,kb, 4\,b of local history are used for the final design.

One more interesting insight worth mentioning is that the HD base predictor performs almost strictly better than a typical bimodal predictor on the synthetic traces. This can be seen in \autoref{BM-Compare} where most of the overlap occurs on the correctly predicted branches of the reference bimodal, while the selected configuration (rightmost in the table) corrects  71\% of the mispredictions, with an overall accuracy to 90.16\%.
\begin{table}[t]
\centering
\begin{tabular}{cccc}
                                                                                                  & \multicolumn{3}{c}{HD Base Predictor Configurations}                                                                                                                                                                                                                                    \\ \hline
\multicolumn{1}{|c|}{Specs}                                                                       & \multicolumn{1}{c|}{\begin{tabular}[c]{@{}c@{}}1\,kb HV\\ 6\,b\\ 1\,k entries\end{tabular}} & \multicolumn{1}{c|}{\begin{tabular}[c]{@{}c@{}}1\,kb HV\\ 3\,b\\ 2\,k entries\end{tabular}} & \multicolumn{1}{c|}{\begin{tabular}[c]{@{}c@{}}1\,kb HV\\ 4\,b\\ 2\,k entries\end{tabular}} \\ \hline
\multicolumn{1}{|c|}{Accuracy}                                                                    & \multicolumn{1}{c|}{91.24\%}                                                               & \multicolumn{1}{c|}{85.37\%}                                                               & \multicolumn{1}{c|}{90.16\%}                                                               \\ \hline
\multicolumn{1}{|c|}{\begin{tabular}[c]{@{}c@{}}Predictions\\ matching\\ Bimodal\end{tabular}} & \multicolumn{1}{c|}{82.04\%}                                                               & \multicolumn{1}{c|}{82.37\%}                                                               & \multicolumn{1}{c|}{82.29\%}                                                               \\ \hline
\multicolumn{1}{|c|}{\begin{tabular}[c]{@{}c@{}}Both\\ mispredict\end{tabular}}                & \multicolumn{1}{c|}{5.08\%}                                                               & \multicolumn{1}{c|}{8.82\%}                                                               & \multicolumn{1}{c|}{5.87\%}                                                                \\ \hline
\multicolumn{1}{|c|}{\begin{tabular}[c]{@{}c@{}}Only HD\\ predicted\\ correct\end{tabular}}    & \multicolumn{1}{c|}{74.53\%}                                                               & \multicolumn{1}{c|}{58.33\%}                                                               & \multicolumn{1}{c|}{71.81\%}                                                               \\ \hline
\multicolumn{4}{l}{Bimodal: 8\,kb+2\,kb, hysteresis 82.44\% accuracy}                                                                                                                                                                                                                                                                                                                     
\end{tabular}
\caption{Detailed comparison between 10\,kb bimodal and 3 viable configurations. The HD predictors have almost no overlap in mispredictions, making them a strictly better option as a base predictor.}
\label{BM-Compare}
\end{table}

\subsection{HYPRE predictor configuration}
With the base predictor configured the whole HYPRE predictor can be now outlined. The exact specifications of the predictor are shown in  \autoref{HD-Table}. The table has two configurations an ideal, and a realistic version of the predictor. As the hypervectors are much larger than typical vectors used in current processors, the simulation of the above design is very slow even when using traces, especially when compared to the other prominent designs like TAGE or Perceptron. As such studies are performed on the aforementioned synthetic traces which aim at stressing the entire design and on a very small subset of Championship Branch Prediction (CBP)\cite{5thPrediction} traces that produce results in a reasonable amount of time that permit tuning of the design.

\begin{figure*}[!t]
    \centering
    \includegraphics[width=\linewidth]{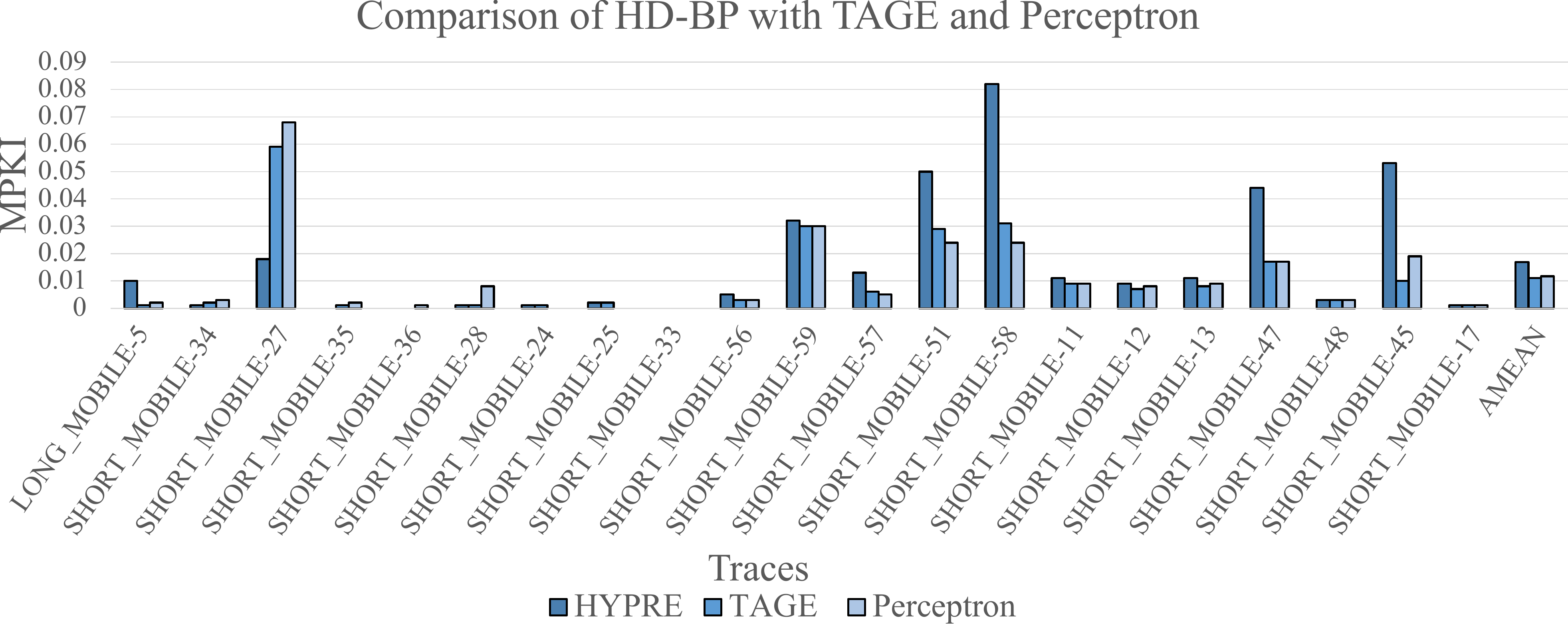}
    \caption{Comparison of HYPRE with TAGE and Perceptron on a limited set of CBP traces.}
    \label{HDTGPC}
\end{figure*}

The ideal design is used for the experiments mentioned above as an over-provisioned version that allows more introspection into the behavior of the predictor. The realistic design differs in the length of the hypervectors that are responsible for the predictions. In particular for history lengths 8--256 the length is 1024 and the long history predictions are 4096. As shown in \autoref{sensei} as long as the vector length is longer than a certain length (and the comparison threshold is adjusted accordingly) the performance of the SDM is not affected. The longer histories are kept at 4096 to avoid branches rotating more than the length of the HV.

Additionally, the ideal version uses both Taken and Not Taken outcome vectors. The use of both these vectors in the ideal version has negligible effect in accuracy, and is mostly used for introspection purposes. Specifically this helps find branch patterns that are stored in both the T and the NT vectors of a specific history length and can be useful in identifying other features that can be used to differentiate the cases that history length alone does not capture.

Finally the ideal version uses a path history and not a global history that only records the outcome. Through tuning, we have noticed that as a sole feature this change has also negligible effect on the accuracy. It is kept in the ideal version for tuning purposes as it also helps identify weaknesses in the approach. A realistic version most probably would use a standard global history that denotes outcome of each branch with one 1 bit.

\section{Results}
This section presents the results from the synthetic traces and the CBP framework selection \cite{5thPrediction}. In addition, through the results and the experimentation the potential strengths and the  shortcomings of the method are detailed to provide a complete view of how hyper-dimensional branch prediction operates. For the comparison with the TAGE and Perceptron predictors, the latest and most accurate published versions of the them are in their 64\kB size versions \cite{Seznec, Jimenez}. We perform two experiments one using the synthetic branches and another one using the subset of CBP.

\begin{table}[!t]
\centering
\begin{tabular}{c|c|c|l}
\cline{2-4}
 & TAGE & Perceptron & \multicolumn{1}{l|}{HYPRE} \\ \hline
\multicolumn{1}{|c|}{Accuracy} & 98.13\% & 98.45\% & \multicolumn{1}{l|}{98.59\%} \\ \hline
\multicolumn{1}{|c|}{\begin{tabular}[c]{@{}c@{}}Identical\\ predictions\\ w/ HYPRE\end{tabular}} & 98.03\% & 97.89\% &  \\ \cline{1-3}
\multicolumn{1}{|c|}{\begin{tabular}[c]{@{}c@{}}Both\\ mispredict \%\end{tabular}} & 0.61\% & 0.38\% &  \\ \cline{1-3}
\multicolumn{1}{|c|}{\begin{tabular}[c]{@{}c@{}}Only HD\\ predicted\\ correct \%\end{tabular}} & 60.00\% & 53.33\% &  \\ \cline{1-3}
\end{tabular}
\caption{Comparison of HYPRE with TAGE and Perceptron on the synthetic traces.}
\label{CMPSNS}
\end{table}

\subsection{Synthetic Branches}
The synthetic branch results are shown in \autoref{CMPSNS}. From the table we see that all three predictors perform almost identically, with less than 1\% accuracy difference. The HYPRE design is slightly more accurate, mainly because it learns the branch patterns slightly faster than the other predictors. HYPRE therefore, ends up having a few more correct predictions early on when each pattern first emerges. This can also be seen in the last row of  \autoref{CMPSNS} where we observe that for instance, of the branches that perceptron and HD-BP have opposite predictions, split of predicting correct is 46.6\% and 53.3\% respectively.

\subsection{Small CBP traces}
As mentioned earlier, due to the fact that the HYPRE design is much slower to simulate, only results from some of the smallest traces were practical to tune the predictor and run. The results from this limited experiment are presented in \autoref{HDTGPC}, where we see that in real traces, while the results are still close there is a gap in accuracy between HYPRE and the currently used designs. The reported arithmetic mean reported by the CBP framework for this subset is 0.017, 0.011, and 0.012 mispredictions per kilo-instructions for HYPRE, TAGE and Perceptron respectively. Looking at individual workloads, even in this subset, we observe that HYPRE rarely outperforms TAGE and perceptron. 

From the experimentation and the tuning of HYPRE, we can attain several noteworthy points. First, while the results do not outperform TAGE and Perceptron, the prototype shows that it can learn the patterns well enough to compete. This is especially important when taking into account that current designs have undergone various optimizations (i.e. the loop predictor and statistical corrector in TAGE - the additonal features in Perceptron) and careful tuning to handle their own pathological cases. As the addition of features is very easy for HDC, we expect the accuracy to improve with similar additions.
\section{Related Work}
The field of branch prediction has evolved dramatically over the past four decades. Initial designs that used only static predictions, such as \cite{Evers1996}, \cite{Kampe} and  \cite{Chang1996}, but are significantly less accurate rarely used today. The initial dynamic branch predictor designs \cite{Lee1984,Mcfarling1993,Chih-ChiehLee2002,Smith2003,Yeh1991, Yeh2004a,Yeh2004b,Young2004} introduced concepts of local and global history and the use of combination of subpredictors, but ultimately also deliver worse accuracy than modern designs as they fail to extract long history patterns. The predictors assessed in this report are the bimodal, the TAGE and the perceptron, which are the designs that are predominantly used today \cite{Mittal2018}. 

All predictors today are assessed based on a variety of metrics which aim at increasing system performance while adhering to necessary constraints. The most important are accuracy, latency, area and security. Accuracy is perhaps the most important requirement  for branch predictors. Designs use MPKI as a metric of the accuracy of each design. For out-of-order cores the misprediction penalty is very high and consequently their predictors need to be extremely accurate. Current state-of-the-art predictors perceptron \cite{Jimenez} and TAGE \cite{Seznec} experience as few as three to four MPKI on average. While these predictors are highly accurate, recent studies have found that there is still room for improvement as there is a lot of performance that could be gained by capturing branches that are classified as hard to predict \cite{Lin2019,Chauhan2020,Ozturk2010}.

Latency constraints are important as the branch predictor lies on the critical path and needs to provide a prediction almost as soon as the branch address is known. Usually the latency is 1 cycle however, complex designs with slow wires operating at a high clock frequency can make the delay larger \cite{Seznec2005,Jimenez2002,Jimenez2000}. To address high latency some designs also provide less accurate preliminary predictions called micro- and nano- predictions \cite{Loh, Jimenezc}. Area also is a limiting factor for branch predictor as the storage needs of sophisticated designs can be in the order of tens of kilobytes (e.g., 64\kB) as seen in the CBP framework \cite{5thPrediction}.

Security has become an active field of research after practical threats, commonly called Spectre-type attacks were identified \cite{Kocher,Canella2019,Evtyushkin2018,Weisse2018,Koruyeh2018}, that can compromise the system using the branch predictor. Mitigation techniques for the security aspects are not simple and usually come at the expense of performance loss \cite{Vougioukas2019, Kiriansky2018,Yan2018,Qureshi2018,McIlroy2019,Khasawneh2018}.

The above observations that current designs leave performance on the table, and that current structures do not have mechanisms to protect against side-channel attacks without performance loss stems our motivation to investigate an alternative way to produce predictions. To the best of our knowledge, our proposal is the first approach that uses sparse distributed memory and hyper-dimensional computing principles to perform branch prediction.

The theory of sparse distributed memory has grown since its inception \cite{kanerva1988sparse,10.5555/183370.183373}. The field was slow in growing in the start because it was difficult to build a meaningful SDM that solves any practical application. In the more recent years however, with the vast improvements in computing and the amount of complexity designs can feature the field has experienced a recent reemergence\cite{Kanerva2009,Rasanen2016}. Recent studies have shown that SDM principles can be leveraged to solve complex problems, a field commonly referred to as hyper-dimensional Computing\cite{Thomas2020}. The ability to perform pattern matching and sequence prediction has made the method popular for artificial intelligence problems\cite{Imani2020}, especially in cases where neural networks are not as practical. In fact, numerous studies show that using HDC can achieve similar level of accuracy (90+\%) as other types of classifiers for a tiny fraction of the size (in some cases as much as 54x reduction) \cite{Rahimi2017,Imani2020,Kim2020}. 

Another aspect that makes HDC a desirable approach for learning and, in the case of this work, sequence prediction is that it has the ability to learn much faster than traditional ML approaches\cite{Rahimi2016a, Burrello2018}. These studies show that it can learn much faster than support vector machine classifiers, deep neural networks. One noteworthy attribute that other HDC works have focused on is that they are inherently robust and secure\cite{Imani2019a}, and as a consequence can be used for sensitive or critical data. This is point is further strengthened by the inherent ability of SDMs to be able to operate in the presence of noise without loss of performance \cite{kanerva1988sparse,Thomas2020}. Finally some studies have focus on how to implement HDC accelerators in hardware using emerging memory technologies and trends to do in-memory computing directly at the vectors \cite{Rahimi2017,Imani2019,Imani2020,Karunaratne2019}.

\section{Conclusion}

For this study, we have focused on the fact that current branch predictor designs while very accurate, can still improve. Hard to predict branches, robustness from noise and interference, low latency and size are all areas where improvements can be made. Historically, such improvements have come from paradigm shifts in the underlying mechanism that is used to generate the predictions. 

In the field of artificial intelligence hyper-dimensional computing has demonstrated that it can be used to deliver high accuracy, inherently robust, fast classifications and predictions with minimal budget. All these attributes match perfectly with the need for advancement in the branch prediction.

This motivated us to investigate how to perform branch prediction using hyper-dimensional computing. We offer a deep introspective look into the mechanisms that make HDC work so efficiently and show how it can be adapted to encode branch patterns, store and identify them and ultimately deliver accurate predictions that are hard to tamper with.

To prove the feasibility of the theoretical model we detail a novel branch predictor prototype called HYPRE, and test it on a limited set of synthetic and real traces. The results show that the HYPRE predictor can learn branch patterns very fast and achieve high accuracy, comparable to realistic designs and deliver the predictions in time through the proposed analog comparison. The vectors which predictions are stored in are also extremely tolerant to bit flips making them inherently hardy against side-channel attacks that try to poison the predictor state. 

With respect to the sensitivity the vectors, we showed that 1024 bits are enough to deliver reliable behavior that retains high accuracy. Keeping the vectors in this length can lead to significant savings as they perform a similar role to the TAGE tables, but are much smaller in size. While the vector length is sufficient, the magnitude of the vectors that store the predictions can possibly saturate and be more of an issue especially for very long executions. 

Improving the accuracy of the state of the art was not the goal of this work as we focused on proving that Hyper-dimensional computing is worth considering as a viable design choice for branch prediction, either as a standalone or as part of a predictor that features current designs as well. However, we believe that with future optimizations and tuning, our designs has the capability to improve accuracy as well.

In terms of future work, we aim to expand the number of traces and workload we test the prototype on. Additionally we plan to investigate how to incorporate additional features, similar to the ones used by Perceptron. Finally, we aim to explore the Hybrid designs that merge HDC branch prediction with more traditional approaches.



\bibliographystyle{IEEEtranS}
\bibliography{main}

\end{document}